\documentclass[a4paper]{article}
\usepackage[pdftex]{graphicx}
\usepackage{INTERSPEECH2021}

\title{DiCOVA-Net: Diagnosing COVID-19 using Acoustics based on Deep Residual Network for the DiCOVA Challenge 2021}

\name{Jiangeng Chang$^1$, Shaoze Cui$^{1,2}$, Mengling Feng$^{1,3}$}

\address{
  $^1$Saw Swee Hock School of Public Health, Natinal University Health System,\\ National University of Singapore, Singapore\\
  $^2$Institute of Information and Decision Technology, Dalian University of Technology, Dalian, China\\
  $^3$Institute of Data Science, National University of Singapore, Singapore
  }

\email{ephjia@nus.edu.sg, e0700415@u.nus.edu, ephfm@nus.edu.sg}

\begin{document}

\maketitle
\begin{abstract}
  In this paper, we propose the DiCOVA-Net, a deep residual network-based method for identifying COVID-19 infected patients using acoustic recordings of their coughs. This classification problem is complicated by the fact that there are far more healthy people than infected patients. We use data augmentation and cost-sensitive methods to improve the model's ability to recognize the minority class (infected patients). In addition, due to the unique nature of this task, we use some fine-tuning techniques to fine-tune the pre-training ResNet50. We also use ensemble learning to integrate prediction results from multiple base classifiers generated using different random seeds to improve the model's generalizability. We ran tests with the DiCOVA challenge dataset to see how well the proposed DiCOVA-Net performed. The results show that our method has achieved 85.43\% in AUC, among the top of all competing teams.
  
\end{abstract}

\section{Introduction}

Since the COVID-19 outbreak in 2020, countries all over the world have been shrouded in an eerie haze. Until recently, people in many countries were at risk of infection. Countries are actively working to avoid large-scale infection in their countries as part of their response to the epidemic. China, the United States, and the United Kingdom have all developed COVID-19 vaccines and have urged people to get vaccinated. However, the vaccine's protection period is limited, and many people are hesitant to be immunized. As a result, effective testing is still required to detect COVID-19 infected individuals.

X-rays and computed tomography were the primary COVID-19 tests during the outbreak's early stages (CT). When medical images of the lungs show symptoms such as shadows, the patient is more likely to have COVID-19, according to \cite{Xie2020}. This detection method produced relatively good results at the start of the epidemic and played an important role in the epidemic's spread. With the advancement of COVID-19 research, the swab test has become a reliable method of detecting COVID-19 infection. Currently, most countries use a throat swab or nose swab test. Furthermore, some studies have shown that saliva may be an effective test method. \cite{Teo2021}.
Other testing methods are being investigated in addition to the methods for infected people mentioned above. Numerous studies have shown that COVID-19 primarily damages lung tissue and has an effect on the respiratory system, according to \cite{Grieco2020}. As a result, it is hypothesized that COVID-19 could cause significant changes in the acoustic characteristics of infected patients. We will investigate a non-invasive method for detecting COVID-19 infected individuals in this paper by using cough audio data to build a classification model.

Machine learning and deep learning have made remarkable advances in many fields in recent years, including image processing \cite{DeCarvalhoFilho2018}, speech recognition \cite{Wu2018}, and text analysis \cite{Sun2017}. In the aftermath of the COVID-19 outbreak, researchers are also attempting to use artificial intelligence technology to combat the epidemic. Tuli et al. \cite{Tuli2020}, for example, created a model based on cloud computing and machine learning to forecast the growth and trend of the COVID-19 pandemic. Yan et al. \cite{Yan2020} developed an XGBoost machine learning-based model capable of predicting the mortality rates of patients who had been hospitalized for more than 10 days Ardakani et al. \cite{Ardakani2020} used CT images to manage COVID-19 in routine clinical practice. Based on these findings, this study aims to identify COVID-19 infected patients using audio recordings of their coughing.

We used various data augmentation, transfer learning, cost-sensitive learning, and ensemble learning techniques to improve the recognition ability of our model, dubbed the DiCOVA-Net. We use a novel randomness method to further improve the model's robustness, and the ensemble results outperform the previous methods.

The rest of the paper is structured as follows. The second section looks at related work. The proposed method is described in detail in Section 3. The experiment and its results are presented in Section 4. This research is summarized in the last section.

\section{Related Work}

In this paper, the task of identifying COVID-19 infected people using acoustic data falls under the category of speech recognition. In the early stages of speech recognition research, scholars used an acoustic model based on Gaussian Mixture Models (GMMs) to achieve a certain recognition effect. Scholars discovered that deep neural networks (DNNs) perform significantly better than GMMs as a result of recent advancements in deep learning methods, \cite{Deng2013}. A spectrogram is a visual representation of a signal's frequency spectrum as it changes over time, which aids deep learning models in modeling audio data. \cite{Sharma2020}.

The problem of imbalanced data is common in medical practice. We discovered, through analysis, that the identification of COVID-19 infected people suffers from the same problem of imbalanced data. The presence of imbalanced data will cause problems in model training, as the model will prioritize recognition of majority samples, which is not what medical staff or managers want. To address the imbalance, researchers have devised a number of strategies. For example, Jiang et al. \cite{Jiang2020} used the data augmentation technique to add noise to the data, with promising results. Furthermore, in addition to processing the data to address the issue of imbalanced data, the loss function of the model can be modified in a cost-effective manner to achieve the goal of modifying the model. Lin et al. \cite{Lin2017}, for example, proposed a new loss function called Focal Loss as an extension of the conventional Cross-Entropy (CE) loss function in the dense object detection task. Experiments confirmed that the Focal Loss function aids in the detection of minority samples in imbalanced data.

Popular deep learning models in image processing and recognition include AlexNet \cite{Krizhevsky2012}, VGG \cite{Simonyan2015}, and ResNet \cite{He2016}, among others. Among the various deep learning models, ResNet is one of the most widely used and well-known. As a result, our proposed model is based on the ResNet50 network. With the advancement of deep learning technology, researchers discovered that it is inconvenient to retrain the entire deep neural network every time they encounter a new task, which takes a significant amount of time and computing resources. As a result, the researchers devised a technique known as "pre-training" \cite{Hendrycks2019}. An already trained deep learning model can be loaded through pre-training, and then the network weight can be partially adjusted (fine-tuning) for specific tasks in the domain, so as to effectively solve new problems. The "pre-train then tune" paradigm is also the core idea of transfer learning \cite{Guo2019}.

Ensemble learning is another popular machine learning technique. The goal of ensemble learning is to improve the overall model's generalization ability and prediction accuracy by combining the outputs of multiple base learners. Some research has found that combining the outputs of multiple deep neural networks using the ensemble learning method can help to improve prediction accuracy \cite{Pang2019, Yang2018}.
The classical methods for generating the base model for ensemble learning primarily focus on changing the training data distribution or the model's structure. In this paper, we look at how randomness affects the model's diversity.

In this paper, we propose a model for COVID-19 infection recognition based on deep residual networks, which combines the technologies of imbalanced data processing, transfer learning, and ensemble learning.

\section{The Proposed DiCOVA-Net Method}

\begin{figure}[h]
  \centering
  \includegraphics[width=\linewidth]{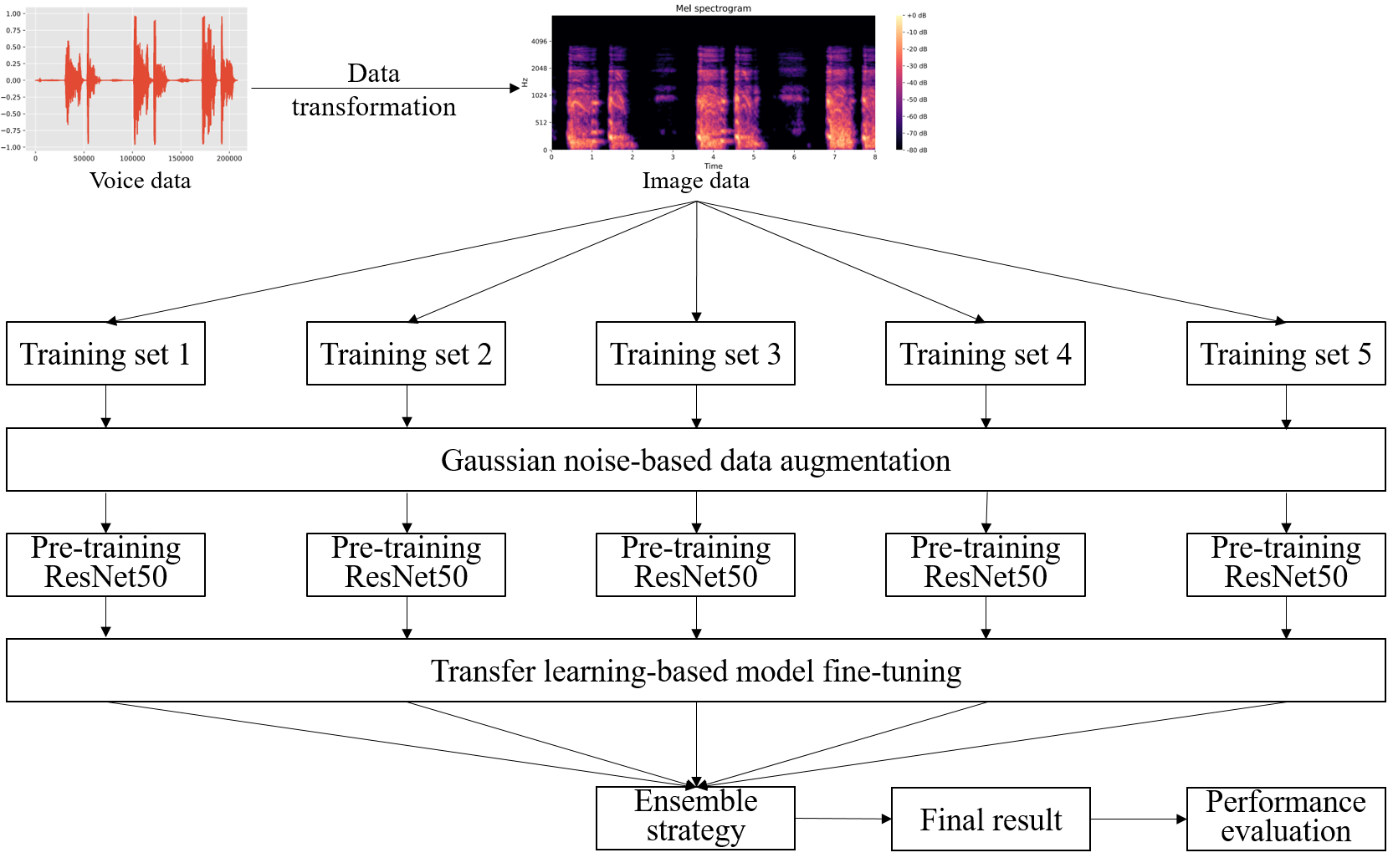}
  \caption{The proposed DiCOVA-Net method.}
  \label{fig:flowchart}
\end{figure}

Figure ~\ref{fig:flowchart} depicts the proposed DiCOVA-Net method. To begin, the input acoustic data must be transformed into spectrogram data. The data is then divided into multiple data subsets for cross validation, and Gaussian noise is added to the minority samples to form new minority samples on each subset. The augmented data set is fed into the ResNet50 pre-training model for fine-tuning, with focal loss serving as the loss function. Finally, when forming the final prediction results, ensemble learning methods are used to combine the outputs of the various models generated by different random seeds.

\subsection{Gussian Noise-based Data Augmentation}

To increase the diversity of the training datasets, data augmentation is frequently required. Furthermore, according to \cite{Jiang2020}, data augmentation can reduce the domain mismatch between the enrolled and test set data. The most common data augmentation methods are flip, rotation, scale, crop, translation, and so on, while we use Gaussian noise. A pseudorandom number generator can generate Gaussian noise, which has a mean of zero and a standard deviation of one. Because DNNs have a large number of parameters, we added Gaussian noise to the data of minority samples (namely COVID-19 infected people) to generate some new synthetic minority samples, which is very beneficial for DNNs and helps to reduce the incidence of overfitting.

\subsection{Deep Residual Network with Transfer Learning}

Residual neural network \cite{He2016} is a traditional method that performed best in the ILSVRC 2015 classification task. A new convolutional neural network structure is the residual neural network. With the success of the VGG network, the depth of the neural network has gained increasing attention, but it is accompanied by the difficult-to-solve problem of gradient disappearance. The residual neural network uses the residual block to solve this problem very well. The residual block can be calculated as Eq.(\ref{eq1}):

\begin{equation}
  y_{r} = h(x_{r})+F(x_{r},W_{r})
  \label{eq1}
\end{equation}

\noindent where $y_r$ represents the label of sample $r$, and $h(x_{r})$ is the prediction that neural network output at the step $r$. The $F(x_{r},W_{r})$ is the residual result of the step $l$.

In the residual neural network, there are two hypotheses: (1) $h(x)$ is the direct mapping. (2) $F(x,W)$ is the direct mapping. Then, the loss $\epsilon$ gradient between the beginning layer and layer $l$ can calculated as Eq.(\ref{eq2}):

\begin{equation}
  \frac{\partial{\epsilon}}{x_{0}} =\frac{\partial{\epsilon}}{x_{l}}(1+\alpha)
  \label{eq2}
\end{equation}

\noindent where $\alpha$ can be presented as Eg.(\ref{eq3}):
\begin{equation}
  \alpha =\frac{\partial{F}}{x_{0}}
  \label{eq3}
\end{equation}

\noindent where $F$ can be presented as Eg.(\ref{eq4}):
\begin{equation}
  F = \sum_{i=0}^{l-1}{f(x_{i},W_{i})}
  \label{eq4}
\end{equation}

Throughout the training process, $\alpha$ cannot always be $-1$, which means that the problem of gradient vanishing will not occur in the residual network

By using a model that is trained on annotated source domain data to predict on unannotated target domain data, transfer learning \cite{transfer} reduces the expense of human annotation. In the few data classification \cite{few-shot}, transfer learning has been shown to be effective. First, we pre-train the ResNet50 on the ImageNet \cite{image}, which has 12 subtrees with 5247 synsets and 3.2 million pictures in total. Then we use the DiCOVA dataset to fine-tune the pre-training neural network structure. We discovered that the model will reach its best point in only a few epoches with the help of pre-training.

\subsection{Focal Loss}
The loss function is used in machine learning and deep learning models to assess the degree of discrepancy between the predicted and real values. If the model forecast is incorrect, the loss function value will be greater. In general, the better the loss function is developed, the better the model's performance. In DNNs, the loss function acts as a "supervisor," guiding model training to progress in the direction of minimizing the loss function in order to locate the network parameter combination that minimizes the loss function. When DNNs are trained, cross entropy (CE), which is derived as Eq.(\ref{eq5}), is frequently employed as the loss function. CE can better represent the differences between various models than the classification error rate and the mean square error (MSE). CE also has the feature of being a convex function, which allows it to discover the global best value when calculating the derivative.

\begin{equation}
  L_{CE} = -\sum_{i=1}^{m}{y_s\cdot{log(p_s)}}
  \label{eq5}
\end{equation}

\noindent where $y_s$ is the label of sample $s$ and $p_s$ is the chance that sample $i$ would be classified as positive.

Cross entropy can produce decent results when the number of samples in each class is small, but it loses effectiveness when the data is uneven. Lin et al. \cite{Lin2017} introduced a novel loss function called focused loss, which is derived as Eq. (\ref{eq6}) to address the problem of imbalanced data.

\begin{equation}
  L_{FL} = -\sum_{r=1}^{m}{\alpha_r(1-p_r)^\gamma {log(p_r)}}
  \label{eq6}
\end{equation}

\noindent where $\gamma$ $(\gamma \geq 0)$ is the focusing parameter used to change the weight of tough and easy samples, and $(1-p r)^\gamma$ is referred to as the modulating factor. Furthermore, $\alpha_r$ is utilized to balance the weights of positive and negative samples.


\subsection{Ensemble Learning}

An ensemble framework aggregates the predictions of multiple base models to get better prediction. Formally, suppose that we have $K$ base models with predictions $\bm{y}_{t + 1} = [y_{t + 1}^1, \cdots, y_{t + 1}^K]$, ensemble is an aggregation function $y_{t + 1} = g(\bm{y}_{t + 1}; \bm{\Phi})$, which has various implementations such as \textit{voting}, \textit{averaging}, and \textit{stacking}~\cite{zhou2012ensemble}. We used the averaging method to ensemble different the same model trained on different distributed feature space.

\subsection{Randomness}
In this paper, we suggested a unique and robust technique for achieving excellent performance on this difficult dataset. Deep learning is rife with randomness, which adds uncertainty to the process. However, the researchers investigated randomization in the weight initialization for deep learning models, which aids in the generation of realized pictures without the need of training \cite{random}. Other scholars investigated the randomness in the layers and models selected for the ensemble methods \cite{multi}. On several domain datasets, this approach produces cutting-edge outcomes. However, none of the preceding techniques investigated the unpredictability in the incoming data. With our understanding, we may conclude that the neural network prioritizes the input entered at the start. So the different order of input batches will make the model start evolving in different places. Therefore, we find the different order of input batches in this challenge data will surely improve the final performance through the experiment. With the infinite search space is expensive, we suggested trying four or five different random seeds and ensemble results together. The novel method can be summarized in $\bm{f}_{t + 1} = [f_{t + 1}^1, \cdots, f_{t + 1}^K]$, the $[f_{t + 1}^1=W(x^r), r\in random(300,2000) $. The $[f_{t + 1}^1, \cdots, f_{t + 1}^K]$ are all based on the same model but with different data input order. The experimental result shows our method improves the model's accuracy and robustness.

\section{Experiments and Results}
\subsection{Datasets}
In the DiCOVA 2021 challenge, \cite{DiCOVA}, we tested our proposed approach. For a two-class classification, this challenge presents a dataset of sound recordings taken from COVID-19 infected and non-COVID-19 individuals. This dataset has a total of 1040 samples, with 965 non-COVID-19 samples and 75 COVID-19 infected samples, resulting in a 13:1 skewed ratio, indicating that this is a highly imbalanced dataset. The average recording duration across subjects is 4.72 seconds (standard error (\textit{S.E.}) $\pm$ 0.07). Furthermore, the challenge organizers use a blind test set with 233 samples, so the results obtained in this paper on the test set are provided by the challenge organizers.

\subsection{Experimental Settings}
To be impartial, the organizers of the competition have previously divided the dataset into a train set and a validation set. The train set contains 822 data points, 772 of which are non-COVID-19 and 50 of which are COVID-19. The validation set contains 218 data points, 193 of which are non-COVID-19 and 25 of which are COVID-19. The organizers additionally provide a five-fold data set to assist participants in gaining a more generic and diversified model.

We use the librosa\footnote{https://github.com/librosa/librosa} to transform the audio data into mel-spectrogram. In the guassian augmentation progress, we use the skimage\footnote{https://github.com/scikit-image/scikit-image} and set the random seed between 0 and 5. We list other parameters in the Table~\ref{table1}.

In the ResNet50, we use the \textit{focal loss} as the loss function and use the \textit{adam} as the optimization method. In model selection, we use the area under the curve (AUC) as an evaluation indicator in each folds.

In the randomness experiment, we only change the random seeds at the beginning of the pipeline.

\begin{table}[t]
  \caption{Main parameters}
  \label{table1}
  \centering
  \begin{tabular}{cc}
    \toprule
    Parameter name      & Settings                      \\
    \midrule
    Sampling Rate            & 2048                                        \\
    Audio length             & 4 sec                                       \\
    FFT window               & 2048                                        \\
    Frame shift              & 512                                         \\
    Epoch                    & 20                                          \\
    Batch size               & 16                                          \\
    Learning rate            & 0.0002                                      \\
    Focal loss $\alpha$      & 0.25                                        \\
    Focal loss $\gamma$      & 2                                           \\
    Random seeds (ensemble)    & 1001,500,1500,2000                        \\
    \bottomrule
  \end{tabular}
\end{table}

\subsection{Results and Discussions}

For this challenge, the organizers present the performance of three baseline methods: Random Forest (RF), Multi-layer Perceptron (MLP), and Logistic Regression (LR). These three methods could not receive audio data, so the organizers use the mel-frequency cepstral coefficients (MFCC) and the delta and delta-delta coefficients methods to extract features. The performance of the three baseline methods which is provided by organizer is shown in Table~\ref{table2} \cite{DiCOVA}. 
According to the results in Table~\ref{table2}, LR has the worst performance, because compared with RF and MLP, LR has the worst nonlinear fitting ability. In addition, the difference between RF's performance on the validation set and the test set is larger than that of MLP, so MLP is more robust.

 \begin{table}[htbp]
  \caption{Baseline performance}
  \label{table2}
  \centering
  \begin{tabular}{ccc}
    \toprule
    Methods & Validation AUC & Test AUC \\
    \midrule
    RF &  70.63 & 67.59 \\
    MLP &  68.81 & 69.91 \\
    LR &  66.97 & 61.97 \\
    \bottomrule
  \end{tabular}
\end{table}

The experiment result of proposed method is shown in Table~\ref{table3}. We compare the performance of fine-tuning model ``Original" with two different loss functions and two different data augmentation methods. The two different loss functions respectively ``cross entropy" (CE) loss and ``focal loss" (FL). The two different data augmentation methods are ``simple duplication" (Dul) and ``Gaussian noise" (Gua). The ``Ensemble" is the performance by integrating four models with different random seeds.

 \begin{table}[htbp]
  \caption{Different methods comparison results}
  \label{table3}
  \centering
  \begin{tabular}{ccc}
    \toprule
    Methods & Validation AUC & Test AUC \\
    \midrule
    CE\_Original &  69.77 & 75.27 \\
    CE\_Dul &  71.15 & 72.92 \\
    CE\_Gua &  71.57 & 75.59 \\
    FL\_Original &  74.17 & 69.78 \\  
    FL\_Dul &  69.92 & 75.10 \\ 
    FL\_Gua &  73.58 & 83.59 \\
    Ensemble & \textbf{76.29} & \textbf{85.43} \\
    \bottomrule
  \end{tabular}
\end{table}

In Table~\ref{table3}, the performance of the fine-tuning model based on ResNet50 is better than that of the traditional machine learning methods such as RF, MLP and LR. In terms of the loss function, CE performs better than FL on the test set, and overfitting occurs when FL is used for training on the original training set. However, after using ``Dul" or ``Gua" for data augmentation, FL performed better than CE on the test set. This proves that a combination of multiple imbalanced data processing methods will achieve better results. In addition, it can also be seen that using Gaussian noise to process imbalanced data is better than directly expanding minority samples. Finally, using randomness to train multiple models for integration can further improve the prediction performance of the entire model.

\section{Conclusions}

In order to use acoustics data to identify patients infected with COVID-19 more accurately, we propose a deep learning method that incorporates multiple image processing techniques. First, we transform the acoustics data into spectrogram data, which can better suit the deep learning model. After that, Gaussian noise-based data augmentation and focal loss are introduced to solve the problem of imbalanced data. Based on the pre-training model of ResNet50, we combine the fine-tuning technology in transfer learning to adjust the weight of the deep neural network to make it more suitable for the identification of COVID-19 infected persons. In addition, in order to make the model we designed more robust, we use ensemble learning to build multiple deep learning models. When training these models, we adopt an advanced data extraction method with randomness and uncertainty to build sample subsets. Our experimental results show that the proposed method can effectively identify the infected with COVID-19 and is superior to other state-of-the-art methods.

\section{Acknowledgements}

We are appreciative to the DiCOVA 2021 Challenge organizers for their efforts in providing participants with data and a platform for the competition. And, this research is supported by the China Scholarship Council (No. 202006060162).

\bibliographystyle{IEEEtran}


\begin{thebibliography}{20}
\bibitem[1]{Xie2020}	X. Xie, Z. Zhong, W. Zhao, C. Zheng, F. Wang, and J. Liu, “Chest CT for Typical Coronavirus Disease 2019 (COVID-19) Pneumonia: Relationship to Negative RT-PCR Testing,” \textit{Radiology}, vol. 296, no. 2, pp. E41–E45, 2020.
\bibitem[2]{Teo2021}	A. K. J. Teo, Y. Choudhury, I. B. Tan, C. Y. Cher, S. H. Chew, Z. Y. Wan, L. T. E. Cheng, L. L. E. Oon, M. H. Tan, K. S. Chan and L. Y. Hsu, “Saliva is more sensitive than nasopharyngeal or nasal swabs for diagnosis of asymptomatic and mild COVID-19 infection,”  \textit{Scientific Reports}, vol. 11, no. 1, pp. 1–8, 2021.
\bibitem[3]{Grieco2020}	D. L. Grieco, F. Bongiovanni, L. Chen, L. S. Menga, S. L. Cutuli, G. Pintaudi, S. Carelli, T. Michi, F. Torrini, G. Lombardi, G. M. Anzellotti, G. D. Pascale, A. Urbani, M. G. Bocci, E. S. Tanzarella, G. Bello, A. M. Dell'anna, S. M. Maggiore, L. Brochard and M. Antonelli, “Respiratory physiology of COVID-19-induced respiratory failure compared to ARDS of other etiologies,”  \textit{Critical Care}, vol. 24, no. 1, pp. 1–11, 2020.
\bibitem[4]{DeCarvalhoFilho2018}	A. O. de Carvalho Filho, A. C. Silva, A. C. de Paiva, R. A. Nunes, and M. Gattass, “Classification of patterns of benignity and malignancy based on CT using topology-based phylogenetic diversity index and convolutional neural network,”  \textit{Pattern Recognition}, vol. 81, pp. 200–212, 2018.
\bibitem[5]{Wu2018}	Y. Wu, H. Mao, and Z. Yi, “Audio classification using attention-augmented convolutional neural network,” \textit{Knowledge-Based Systems}, vol. 161, pp. 90–100, 2018.
\bibitem[6]{Sun2017}	S. Sun, C. Luo, and J. Chen, “A review of natural language processing techniques for opinion mining systems,” \textit{Information Fusion}, vol. 36, pp. 10–25, 2017.
\bibitem[7]{Tuli2020}	S. Tuli, S. Tuli, R. Tuli, and S. S. Gill, “Predicting the Growth and Trend of COVID-19 Pandemic using Machine Learning and Cloud Computing,” \textit{Internet of Things}, p. 100222, 2020.
\bibitem[8]{Yan2020}	L. Yan, H.-T. Zhang, J. Goncalves, Y. Xiao, M. Wang, Y. Guo, C. Sun, X. Tang, L. Jing, M. Zhang, X. Huang, Y. Xiao, H. Cao, Y. Chen, T. Ren, F. Wang, Y. Xiao, S. Huang, X. Tan, N. Huang, B. Jiao, C. Cheng, Y. Zhang, A. Luo, L. Mombaerts, J. Jin, Z. Cao, S. Li, H. Xu and Y. Yuan, “An interpretable mortality prediction model for COVID-19 patients,” \textit{Nature Machine Intelligence}, vol. 2, 2020.
\bibitem[9]{Ardakani2020}	A. A. Ardakani, A. R. Kanafi, U. R. Acharya, N. Khadem, and A. Mohammadi, “Application of deep learning technique to manage COVID-19 in routine clinical practice using CT images: Results of 10 convolutional neural networks,” \textit{Computers in Biology and Medicine}, vol. 121, p. 103795, 2020.
\bibitem[10]{Deng2013}	L. Deng, G. Hinton, and B. Kingsbury, “New types of deep neural network learning for speech recognition and related applications: An overview,” in \textit{2013 IEEE International Conference on Acoustics, Speech and Signal Processing (ICASSP)}, 2013, pp. 8599–8603.
\bibitem[11]{Sharma2020}	J. Sharma, O. Granmo, and M. Goodwin, “Environment Sound Classification using Multiple Feature Channels and Attention based Deep Convolutional Neural Network,” in \textit{INTERPEECH 2020}, 2020, pp. 1186–1190.
\bibitem[12]{Jiang2020}	T. Jiang, M. Zhao, L. Li, and Q. Hong, “The XMUSPEECH system for short-duration speaker verification challenge 2020,” in \textit{INTERSPEECH 2020}, 2020, pp. 736–740.
\bibitem[13]{Lin2017}	T. Y. Lin, P. Goyal, R. Girshick, K. He, and P. Dollár, “Focal loss for dense object detection,” in \textit{Proceedings of the IEEE International Conference on Computer Vision (ICCV)}, 2017, pp. 2980–2988.
\bibitem[14]{Krizhevsky2012}	A. Krizhevsky, I. Sutskever, and G. E. Hinton, “ImageNet Classification with Deep Convolutional Neural Networks,” in \textit{Advances in neural information processing systems (NIPS)}, 2012, pp. 1097–1105.
\bibitem[15]{Simonyan2015}	K. Simonyan and A. Zisserman, “Very deep convolutional networks for large-scale image recognition,” in \textit{3rd International Conference on Learning Representations (ICLR)}, 2015, pp. 1–14.
\bibitem[16]{He2016}	K. He, X. Zhang, S. Ren, and J. Sun, “Deep residual learning for image recognition,” in \textit{Proceedings of the IEEE Conference on Computer Vision and Pattern Recognition (CVPR)}, 2016, pp. 770–778.
\bibitem[17]{Hendrycks2019}	D. Hendrycks, L. Kimin, and M. Mazeika, “Using Pre-Training Can Improve Model Robustness and Uncertainty,” in \textit{Proceedings of the 36th International Conference on Machine Learning (ICML)}, 2019, pp. 2712–2721.
\bibitem[18]{Guo2019}	Y. Guo, H. Shi, A. Kumary, K. Grauman, T. Rosing, and R. Feris, “Spottune: Transfer learning through adaptive fine-tuning,” in \textit{Proceedings of the IEEE Conference on Computer Vision and Pattern Recognition (CVPR)}, 2019, pp. 4805–4814.
\bibitem[19]{Pang2019}	T. Pang, K. Xu, C. Du, N. Chen, and J. Zhu, “Improving adversarial robustness via promoting ensemble diversity,” in \textit{Proceedings of the 36th International Conference on Machine Learning (ICML)}, 2019, pp. 4970–4979.
\bibitem[20]{Yang2018}	X. Yang, Z. Zeng, T. Sing G., L. Wang, V. Chandrasekar, and S. C. H. HOI, “Deep learning for practical image recognition: Case study on kaggle competitions,” in \textit{Proceedings of the ACM SIGKDD International Conference on Knowledge Discovery and Data Mining}, 2018, pp. 923–931.
\bibitem[21]{random}
   K.\ He, Y.\ Wang, J.\ Hopcroft, 
   ``A Powerful Generative Model Using Random Weights for the Deep Image Representation'',
   \textit{Proceedings of the 30th International Conference on Neural Information Processing Systems (NIPS)}, 2016, pp.~631-639.
\bibitem[22]{multi}
   K.\ Kowsari, M.\ Heidarysafa, D.\ E.\ Brown, K.\ J.\ Meimandi, L.\ E.\ Barnes,
   ``RMDL: Random Multimodel Deep Learning for Classification'',
   \textit{ Proceedings of the 2nd International Conference on Information System and Data Mining }, 2018, pp.~19-28.
\bibitem[23]{transfer}  SJ.\ Pan, Q.\ Yang, 
   ``A survey on transfer learning'',
   \textit{ IEEE Transactions on knowledge and data engineering }, vol. 22, no. 10, pp.~1345-1359, 2009.
\bibitem[24]{few-shot}  Q.\ Sun, Y.\ Liu, T.\ Chua, B.\ Schiele,
   ``Meta-Transfer Learning for Few-Shot Learning'',
   \textit{ Proceedings of the IEEE Conference on Computer Vision and Pattern Recognition (CVPR) }, 2019, pp.~1345-1359.
\bibitem[25]{image}  J.\ Deng, W.\ Dong, R.\ Socher, L.\ Li, K.\ Li, F.\ Li,
   ``ImageNet: A large-scale hierarchical image database'',
   \textit{Proceedings of the IEEE Conference on Computer Vision and Pattern Recognition (CVPR)}, 2009, pp.~248-255.
\bibitem[26]{zhou2012ensemble}  Z.\ Zhou,
   ``Ensemble methods: foundations and algorithms'',
   \textit{ Chapman and Hall/CRC press}, 2012.
\bibitem[27]{DiCOVA}
   A.\ Muguli, L.\ Pinto, N.\ R, N.\ Sharma, P.\ Krishnan, P,\ K,\ Ghosh, R.\ Kumar, S.\ Ramoji, S.\ Bhat, S.\ R.\ Chetupalli, S.\ Ganapathy, V.\ Nanda, 
   ``DiCOVA Challenge: Dataset, task, and baseline system for COVID-19 diagnosis using acoustics'',
   \textit{ arXiv preprint ArXiv: 2103.09148}, 2021.
\end{thebibliography}

\end{document}